%% file: main.tex
\documentclass[conference]{IEEEtran}
\IEEEoverridecommandlockouts
\usepackage{cite}
\usepackage{amsmath,amssymb,amsfonts}
\usepackage{algcompatible}
\usepackage{algorithm}
\usepackage{graphicx}
\usepackage{textcomp}
\usepackage{mathtools}   
\usepackage{lipsum} 
\usepackage{subfig}
\usepackage{amsfonts}
\usepackage{amssymb}
\usepackage{color}
\usepackage{caption}
\usepackage{mathtools}
\usepackage{relsize}
\usepackage{amsthm}
\usepackage{amssymb}
\usepackage{amsmath}
\usepackage{algpseudocode}
\usepackage{setspace}
\usepackage{listings}
\usepackage{graphicx}

\usepackage{xcolor}

\newcommand{\tool}{\textsc{ScaleBench}}

\def\BibTeX{{\rm B\kern-.05em{\sc i\kern-.025em b}\kern-.08em
    T\kern-.1667em\lower.7ex\hbox{E}\kern-.125emX}}
\begin{document}
\title{ScaleBench: An Extensible and Reproducible Experimental Platform for Kubernetes Autoscaling Research}
\author{
\IEEEauthorblockN{Ze Wang}
\IEEEauthorblockA{\textit{School of Computing, Engineering,}\\
\textit{and the Built Environment} \\
\textit{Edinburgh Napier University}\\
Edinburgh, UK \\
z.wang3@napier.ac.uk}
\and
\IEEEauthorblockN{Sajid Alam}
\IEEEauthorblockA{\textit{School of Computing, Engineering,}\\
\textit{and the Built Environment} \\
\textit{Edinburgh Napier University}\\
Edinburgh, UK \\
s.alam2@napier.ac.uk}
\and
\IEEEauthorblockN{Amjad Ullah}
\IEEEauthorblockA{\textit{School of Computing, Engineering,}\\
\textit{and the Built Environment} \\
\textit{Edinburgh Napier University}\\
Edinburgh, UK \\
a.ullah@napier.ac.uk}

}

\maketitle
\begin{abstract}
\input{sections/00_abstract}
\end{abstract}
\input{sections/01_intro}

\input{sections/02_related-work}

\input{sections/03_methodology}
\input{sections/04_evaluation}
\input{sections/05_conclusion}
\section*{Acknowledgments}
This work is co-funded by the European Union’s Horizon Europe programme under grant agreement No. 101135012, and by UK Research and Innovation (UKRI) under grant agreement No. 10102651, as part of the project Swarmchestrate: Application-level Swarm-based Orchestration Across the Cloud-to-Edge Continuum.
\bibliographystyle{IEEEtran}
\bibliography{Ref}
\end{document}

%% file: sections/00_abstract.tex
Autoscaling enables cloud-native applications to meet their SLOs by dynamically adjusting computational resources at runtime. Recent research has explored a broad range of advanced autoscaling approaches, including rule-based, optimisation-based, and machine learning-based techniques, to address the limitations of traditional resource management strategies in cloud-native environments. Despite significant advances, a considerable gap remains between academic innovations and their practical adoption in real-world environments. For instance, Kubernetes deployments commonly rely on the reactive, CPU-based Horizontal Pod Autoscaler, limiting the adoption of more advanced autoscaling approaches. Recent studies have investigated challenges such as autoscaler configuration and reproducible performance evaluation; however, the practical deployment, integration complexity, and operational overhead of research-based autoscalers in Kubernetes environments have received limited attention. To address this gap, we present \tool{}, a native Kubernetes experimental platform to enable extensible and reproducible evaluation of autoscaling solutions. Unlike existing approaches, \tool{} integrates autoscalers as native Kubernetes components, allowing researchers to assess not only scaling performance but also deployment characteristics and integration requirements within realistic cloud-native environments. We demonstrate the effectiveness of \tool{} by evaluating multiple autoscaling approaches through a case study. The results highlight its capability to support systematic, repeatable, and practical assessment of Kubernetes autoscaling approaches. 

%% file: sections/01_intro.tex
\section{Introduction} \label{sec:intro}

Cloud-native applications have become the foundation of modern digital services, enabling organisations to rapidly develop and deploy scalable applications~\cite{Carrion22, LC26, ullah2023orchestration}. However, the unpredictable nature of modern workloads creates significant challenges for efficiently provisioning resources while maintaining application performance and cost efficiency. Autoscaling has emerged as a fundamental mechanism for enabling native cloud applications to satisfy their Service Level Objectives (SLOs) by adjusting their resource needs at runtime~\cite{NYK+20}. Hence, it reduces operational complexity and improves resource utilisation compared to static provisioning. Consequently, extensive research has been devoted to developing advanced autoscaling techniques that improve the responsiveness, efficiency, and reliability of resource management in cloud-native environments~\cite{TVK22, JJ25, ullah2018control}.

Despite significant progress in autoscaling research, a substantial gap remains between academic advancements and their practical adoption in real-world cloud environments. Existing autoscaling research has predominantly focused on algorithm design and scaling performance, while comparatively little attention has been paid to the practical deployment and integration of research-based autoscalers in Kubernetes environments.~\cite{SGK+22, SEK+23, XWL+26}. As a result, research prototypes often rely on customised experimental setups, application-specific assumptions, or non-standard evaluation methodologies, making it difficult to reproduce results, compare different approaches, and assess their practical deployment implications. This challenge is particularly evident in Kubernetes~\cite{Carrion22}, where the built-in Horizontal Pod Autoscaler (HPA) remains the most widely adopted autoscaling mechanism despite its reliance on predefined metrics such as CPU utilisation. Consequently, important operational aspects---including Kubernetes-native integration, deployment complexity, controller resource overhead, and portability across different applications and workloads---are rarely evaluated systematically.

The factors mentioned above directly influence the engineering effort and operational feasibility of adopting research-based autoscalers in practice. For example, an autoscaler may achieve superior scaling performance while still introducing significant deployment complexity, requiring application-specific modifications, or incurring substantial runtime overhead. Consequently, deployment feasibility should be evaluated alongside conventional performance metrics when assessing Kubernetes autoscaling approaches, as practical adoption depends not only on scaling effectiveness but also on ease of integration and operational overhead. This highlights the need for an extensible and reproducible experimental platform that enables a comprehensive assessment of both the effectiveness of scaling and the practical characteristics of the deployment. This paper is a step forward in this direction. More specifically, our key contributions are as follows: 
\begin{itemize}
    \item We introduce \tool{}\footnote{{https://github.com/ZeWang42/Swarmchestrate-Decentralised-Auto-Scaling}}, an open-source, modular, autoscaling evaluation platform that treats autoscalers as deployable Kubernetes components and considers both scaling performance and practical deployment characteristics. Furthermore, it combines manifest-based deployment, reusable application \& autoscaler registries, and automated experiment orchestration to provide a complete evaluation workflow. 
    \item A comprehensive evaluation of \tool{} using state-of-the-art autoscalers to demonstrate its effectiveness and usability.
\end{itemize}

The remainder of this paper is structured as follows. Section~\ref{sec:related} reviews the deployment methodologies of existing autoscaling approaches and practical considerations for production environments. Section~\ref{sec:scalebench} presents the design and implementation of \tool{}. Section~\ref{sec:prototype} demonstrates \tool{}'s deployment using representative autoscalers, workloads and benchmark applications. Finally, Section~\ref{sec:Conc} concludes the paper and outlines future work.

%% file: sections/02_related-work.tex
\section{Related Work} \label{sec:related}
The landscape of autoscaling literature encompasses a wide range of scaling strategies and decision-making techniques~\cite{TVK22, DNK23, JJ25, ullah2018control}. However, the focus of this paper is not to review autoscaling techniques themselves, but rather to examine how they are evaluated. We first discuss representative autoscaling solutions from the perspective of their deployment methodology, and then highlight the key challenges affecting their practical applicability and deployment in Kubernetes environments.

\subsection{Deployment Architectures}
Simulation-based evaluation remains a widely adopted approach for assessing autoscaling algorithms, particularly in early studies on reinforcement learning and predictive autoscaling. For example, Zanussi et al.~\cite{ZTL+26} and Ding et al.~\cite{DLC+25} evaluated their proposed scaling policies using CloudSim~\cite{CRB+11} and AutoScaleSim~\cite{ATTG21} simulation frameworks, respectively, without deploying either the target application or the autoscaler within a real Kubernetes or container orchestration environment. While simulation provides a controlled and cost-effective means of exploring algorithmic behaviour, it often abstracts away the deployment, integration, and operational complexities encountered in production cloud-native systems.

Beyond simulation-based evaluation, many recent studies assess autoscaling approaches using real Kubernetes clusters. However, the deployment architectures adopted for these autoscalers vary considerably. A common approach is to implement the autoscaler as a centralised control process that periodically collects monitoring data from Prometheus or the Kubernetes Metrics Server, analyses the current system state, and enforces scaling decisions through the Kubernetes API. Some notable contributions to this list include PBScaler~\cite{XWL+24}, QueueFlower~\cite{CLG+24}, HAB~\cite{TWPY21}, and DeepScaling~\cite{WZL+24}, where the autoscaler is logically separated from the managed application; however, it interacts with the Kubernetes control plane through its APIs. In contrast, SmartHPA~\cite{ATWS24} 
deploys the autoscaler on a separate management node that remotely monitors and scales applications on Kubernetes clusters.

A smaller body of research integrates autoscaling mechanisms more closely with the native Kubernetes ecosystem. For example, ge-kube~\cite{RCPN20} extends Kubernetes by introducing dedicated elasticity and placement managers, with its custom scheduler deployed as a native Kubernetes Pod. Mondal et al.~\cite{MWK+23} packaged their machine learning-based autoscaler using the custom Pod autoscaling framework, while PAHPA~\cite{XYA+26} combines a prediction service with the Kubernetes Event-Driven Autoscaler and the native HPA. In contrast to the above approaches, DAS~\cite{SGL+25}  adopts a decentralised architecture by embedding the autoscaling logic as a lightweight binary within each application container, where it runs as a background process alongside the application.


\subsection{Practical Concerns}
Recent studies have increasingly recognised the gap between research-based autoscalers and their practical adoption. For example, Straesser et al.~\cite{SGK+22} identified key challenges that limit production readiness, including workload characteristics, metric selection, and configuration complexity. Straesser et al.~\cite{SEK+23} proposed a practitioner-oriented evaluation methodology that standardises workload design, experiment execution, and performance analysis to improve evaluation consistency.

Xie et al.~\cite{XWL+26} proposed ScalerEval, an end-to-end benchmarking platform which automates the evaluation workflow and reduces human effort. In contrast to ScalerEval, where the autoscaler runs as an asynchronous thread outside the cluster, ScaleBench provides a flexible, Kubernetes-native deployment model. 
Jeong and Jeong~\cite{JJ25} summarised evaluation workloads, metrics and environment. Dashtbani and Tahvildari~\cite{DT25} showed that benchmark application design and deployment characteristics can significantly influence autoscaler performance. While these recent studies have highlighted practical concerns regarding evaluation and adoption, the practical deployment of research-based autoscalers in Kubernetes---including their deployment model, integration effort, and resource overhead---remains comparatively underexplored.

%% file: sections/03_methodology.tex
\section{\tool{}}\label{sec:scalebench}
This section presents the overall vision, architecture and implementation of \tool{}, highlighting how it enables extensible and reproducible evaluation of autoscaling approaches through a Kubernetes-native deployment model.

\subsection{Design Vision}


The goal of \tool{} design is to provide a Kubernetes-native platform in which both applications and autoscalers are deployed as Kubernetes resources. By adopting such a model, \tool{} enables the practical deployment characteristics of research-based autoscalers, such as integration effort and resource overhead, to be evaluated alongside their scaling performance. Furthermore, \tool{} provides a standardised experimental workflow that ensures consistent deployment, workload execution, and monitoring across experiments, enabling reproducible and fair evaluations. Finally, \tool{} is designed with modularity and extensibility in mind, allowing new autoscaling algorithms, applications, and workloads to be integrated with minimal effort.
\subsection{Architecture}

\tool{} follows a client--server architecture. 
The client is responsible for
defining experiment specifications, generating workloads, and
collecting experimental results. The server is deployed within the target Kubernetes cluster, where it manages the experimental workflow, which involves application deployment, autoscaler deployment, and runtime monitoring. Internally, the server adopts a three-layer architecture consisting of an API layer, an orchestration layer, and a Kubernetes interaction layer. The API layer provides the communication interface between the client and the server. The orchestration layer coordinates the experiment lifecycle and prepares Kubernetes resource definitions for deployment. The Kubernetes interaction layer encapsulates all communication with the Kubernetes cluster through the Kubernetes API. 






\tool{} models both applications and autoscalers as Kubernetes resources. This design intends to match practical Kubernetes deployments and place the autoscaler's control loop close to both the managed application and the Kubernetes control plane, thereby minimising communication latency and making it easy to manage. To support extensibility, \tool{} organises applications and autoscalers using dedicated registries. 
Applications are represented by an application registry together with a manifest directory. The registry stores application-level metadata, including the set of deployments and services belonging to the application, its target namespace, the location of its Kubernetes manifests, and the frontend service through which the workload generator accesses the application. The manifest directory contains the Kubernetes resources required to deploy the application, including its deployment, service, and gateway definitions.

Autoscalers are represented through an autoscaler registry and a corresponding manifest directory. The registry stores metadata for each autoscaler, including the location of its Kubernetes manifest templates and configurable parameters, while the manifest directory contains the Kubernetes resources required for deployment, such as controller deployments, service accounts, and RBAC configurations. Consequently, integrating a new application or autoscaler requires only the addition of the relevant manifest templates and metadata to the registry, without requiring any modifications to the underlying experiment workflow.
\subsection{Implementation}
\tool{} follows a client--server architecture that separates experiment management from autoscaler execution. Accordingly, the platform comprises two principal components: a client responsible for configuring experiments, and a server that manages their execution within Kubernetes. The implementation details of these components are described in the following subsections.
\subsubsection{Server}
The server communicates with the client and orchestrates the experimental environment in Kubernetes through the following layers.
\begin{enumerate}
    \item The \textbf{API Layer} is implemented using FastAPI\footnote{https://fastapi.tiangolo.com/} and exposes REST endpoints for managing the experiment lifecycle. The \textit{setup} endpoint receives an experiment specification, including the target application, autoscaler configuration, and workload metadata. It validates that the experiment specification contains all required fields. After validating the request, it forwards the experiment specification to the orchestration layer. Once the experimental environment has been prepared, it notifies the client to start generating the workloads. The \textit{cleanup} endpoint receives a cleanup request and forwards it to the orchestration layer to terminate the experiment.
    \item The \textbf{Orchestration Layer} coordinates the experiment lifecycle and prepares the resource definitions required by the Kubernetes interaction layer. Upon receiving an experiment setup request, it first resolves the requested application from the application registry and invokes the Kubernetes interaction layer to ensure that the application is deployed. Next, it resets each application microservice to a single replica, thereby establishing a consistent initial state across experiments. Once the application is ready, the orchestration layer resolves the requested autoscaler from the autoscaler registry, followed by loading the corresponding manifest templates and rendering them using experiment-specific parameters, such as the target namespace, application deployments, and autoscaler configuration. The rendering process substitutes these experiment-specific values into the manifest templates, producing Kubernetes resource definitions that are passed to the Kubernetes interaction layer for deployment. The monitoring service is then started, and the gateway endpoint of the deployed application is returned to the client. At the end, during the experiment cleanup, the orchestration layer stops the monitoring service, removes the deployed autoscaler, and restores the application to its initial state by resetting all microservices to a single replica.
    \item The \textbf{Kubernetes Interaction Layer} communicates with the API server through the official Kubernetes Python client to manage deployments, services, and custom resources. The layer receives rendered Kubernetes resource definitions from the orchestration layer and performs the corresponding cluster operations, including creating, updating, and deleting Kubernetes resources, resetting deployment replica counts, monitoring deployment readiness, and retrieving application gateway endpoints. During workload execution, the Kubernetes interaction layer periodically collects metrics about the application, autoscaler and cluster. Resource utilisation metrics, including CPU usage, memory usage, replica counts, and node utilisation, are collected through the Kubernetes Metrics API, while Prometheus~\cite{prometheus} provides the required percentile application latency over the previous minute. All metrics are timestamped and stored as CSV files for subsequent analysis.
\end{enumerate}
\subsubsection{Client}
The client accepts an experiment configuration as input. An experiment configuration defines the application under test, the target Kubernetes namespace, the workload, the experiment duration, the autoscaler configuration, and the monitoring parameters. The autoscaler configuration part identifies the autoscaling algorithm, the target deployments it manages, and algorithm-specific parameters, such as replica bounds or utilisation thresholds.

The client reads the experiment configuration and generates all requested combinations of applications, workloads, and autoscalers as a sequence of independent experiments. For each experiment, it invokes the server's setup API and waits until the deployment is ready and the application gateway is returned. It then uses Locust\footnote{{https://docs.locust.io/en/stable/}} in headless mode to generate workload traffic through the application's gateway. The workload is defined by a CSV file, which lists the desired user count at each time. A standard Locust workload script is used to define application-specific behaviours. After workload execution completes, the client invokes the cleanup API and stores the Locust statistics together with the monitoring data collected by the server.

%% file: sections/04_evaluation.tex
\section{Experimental Workflow and Case Study}\label{sec:prototype} 
This section demonstrates the use of \tool{} by deploying and evaluating state-of-the-art autoscalers on representative benchmark applications under diverse workload scenarios. The examples illustrate the platform's workflow and its ability to support reproducible experimentation.

\begin{figure*}[th!]
    \centering

    \subfloat[Latency CDF.\label{fig:ob_wiki_latency_cdf}]{
        \includegraphics[width=0.33\textwidth]{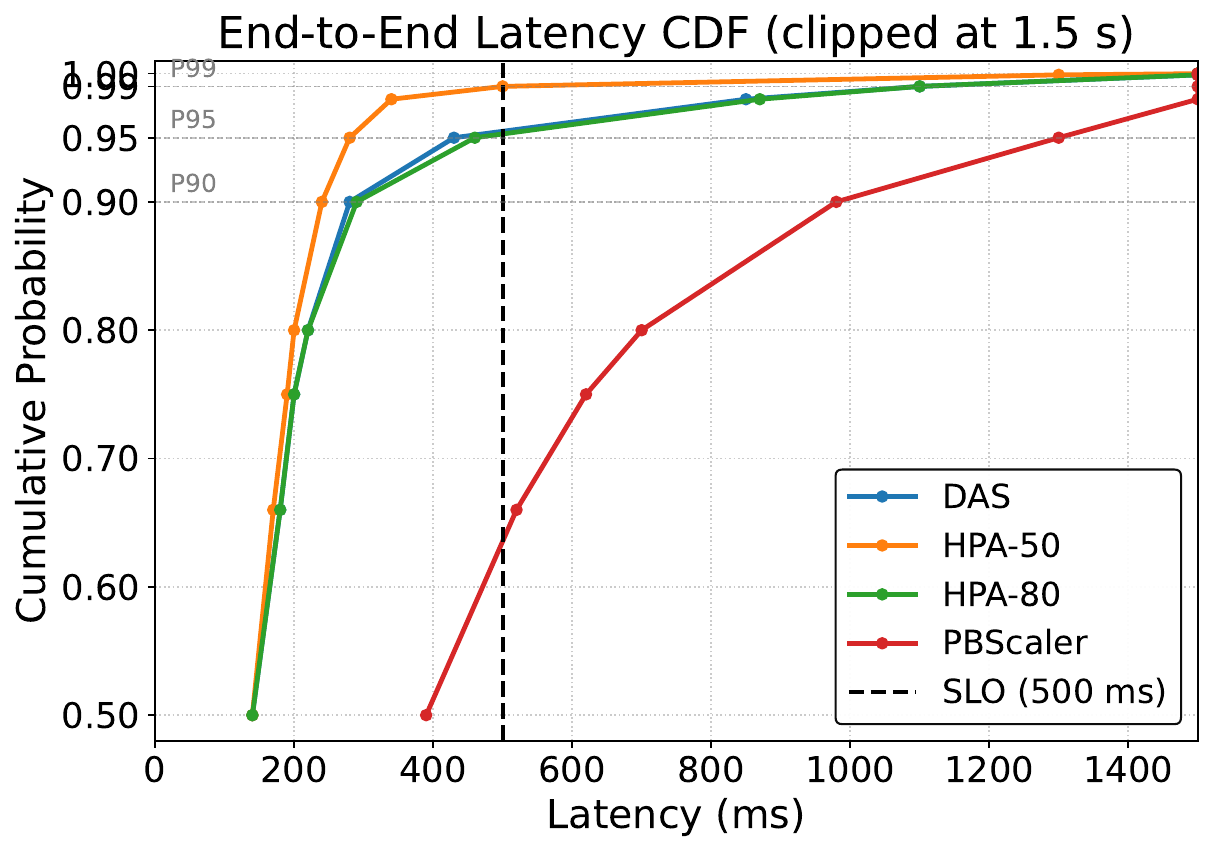}
    }
    \subfloat[SLO violations.\label{fig:ob_wiki_slo_violation_rate}]{
        \includegraphics[width=0.33\textwidth]{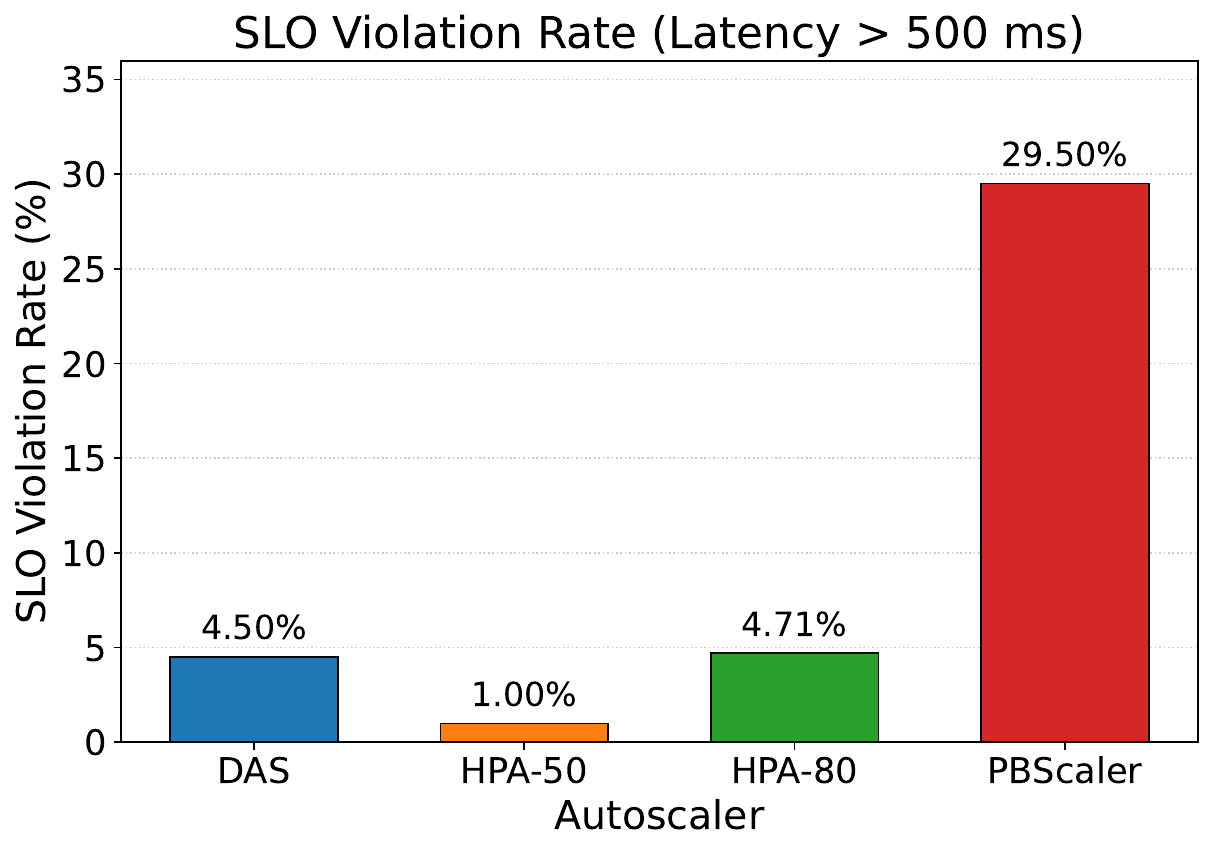}
    }
    \subfloat[CPU and memory usage.\label{fig:ob_wiki_application_resource}]{
        \includegraphics[width=0.33\textwidth]{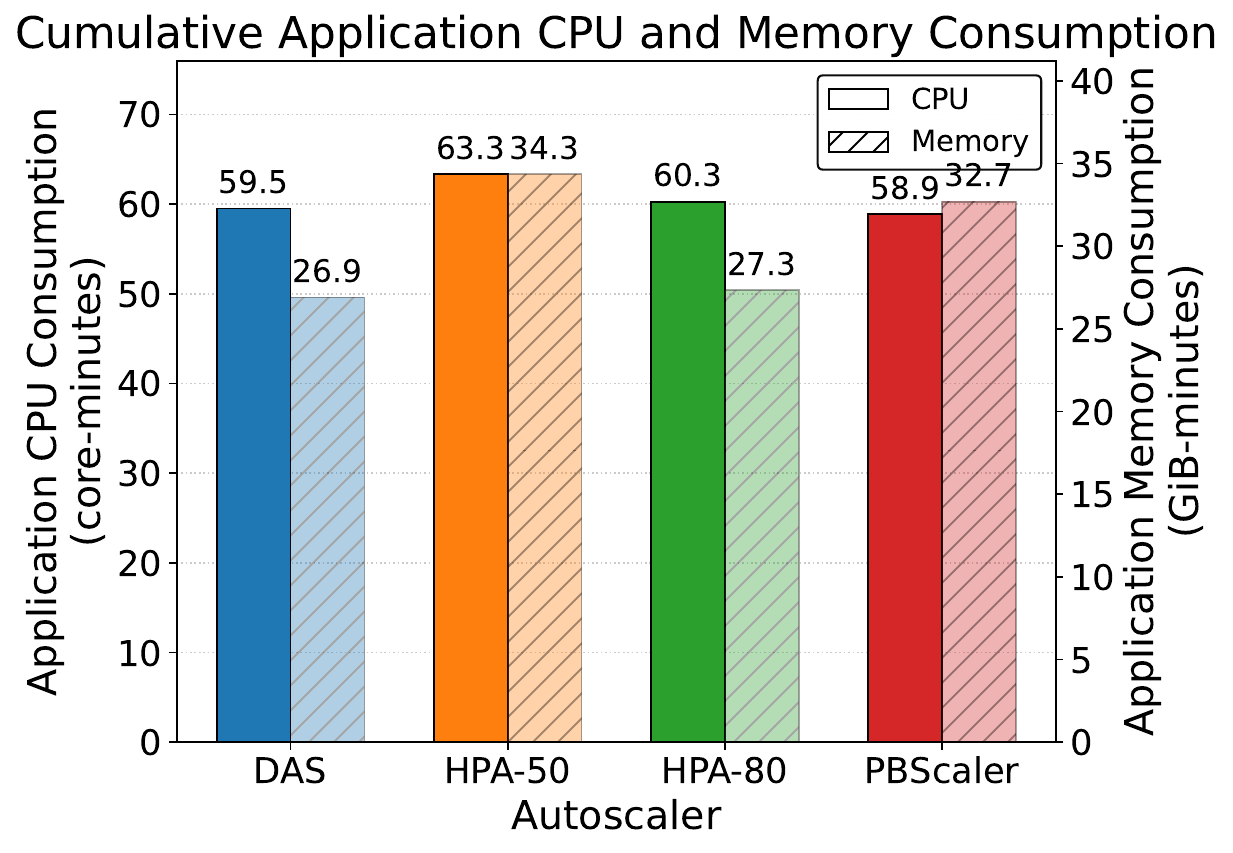}
    }

    \caption{Performance and resource cost of Online Boutique.}
    \label{fig:performance_and_cost}
\end{figure*}
\begin{figure*}[th!]
    \centering

    \subfloat[Application P90 latency.\label{fig:ob_wiki_latency_over_time}]{
        \includegraphics[width=0.495\textwidth]{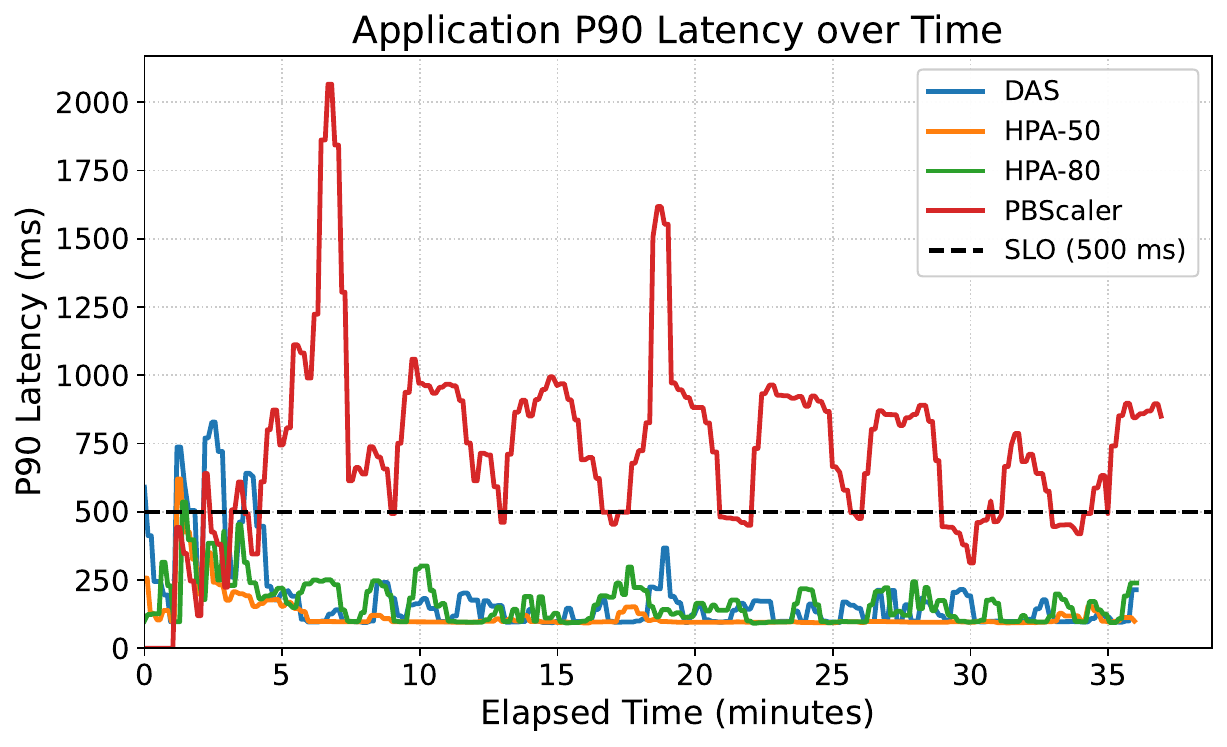}
    }
    \subfloat
    [Average total number of application replicas per minute.\label{fig:ob_wiki_replicas_over_time}]
    {
        \includegraphics[width=0.495\textwidth]{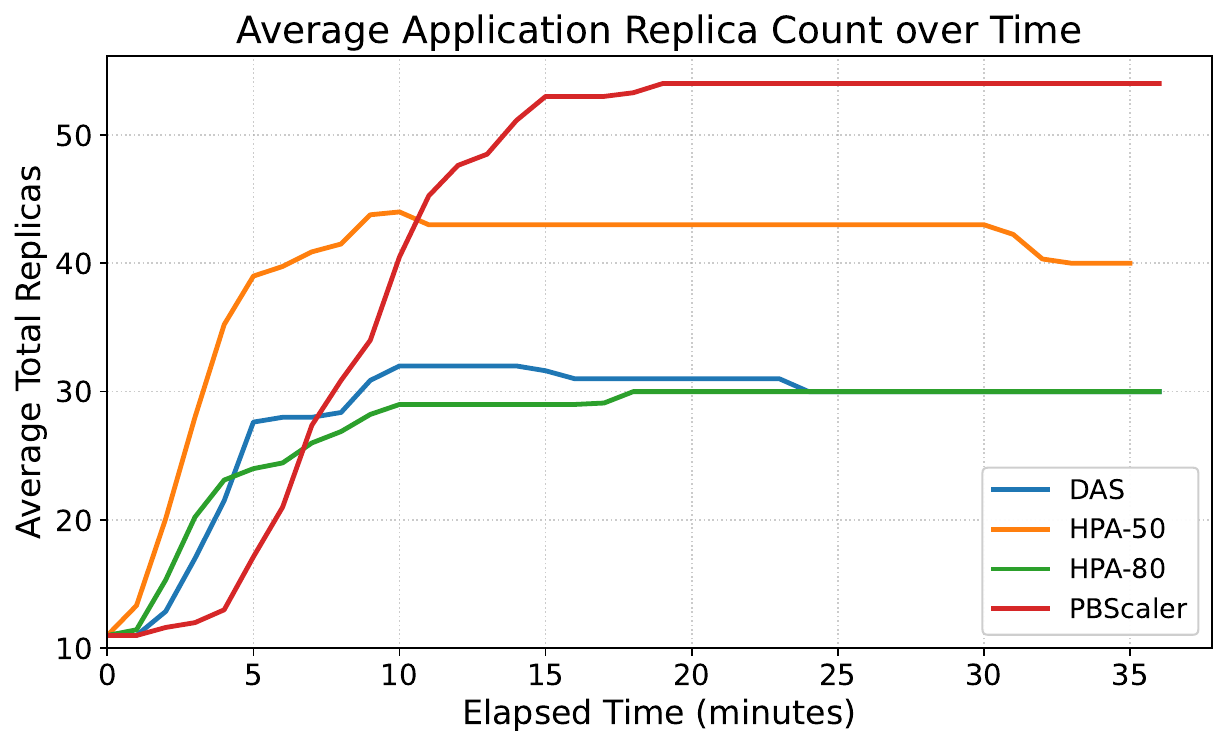}
    }
        \caption{The latency and average replica count of Online Boutique over time.}
    \label{fig:latency_replica_over_time}
\end{figure*}

\begin{figure}[th!]
    \centering

    {
        \includegraphics[width=0.33\textwidth]{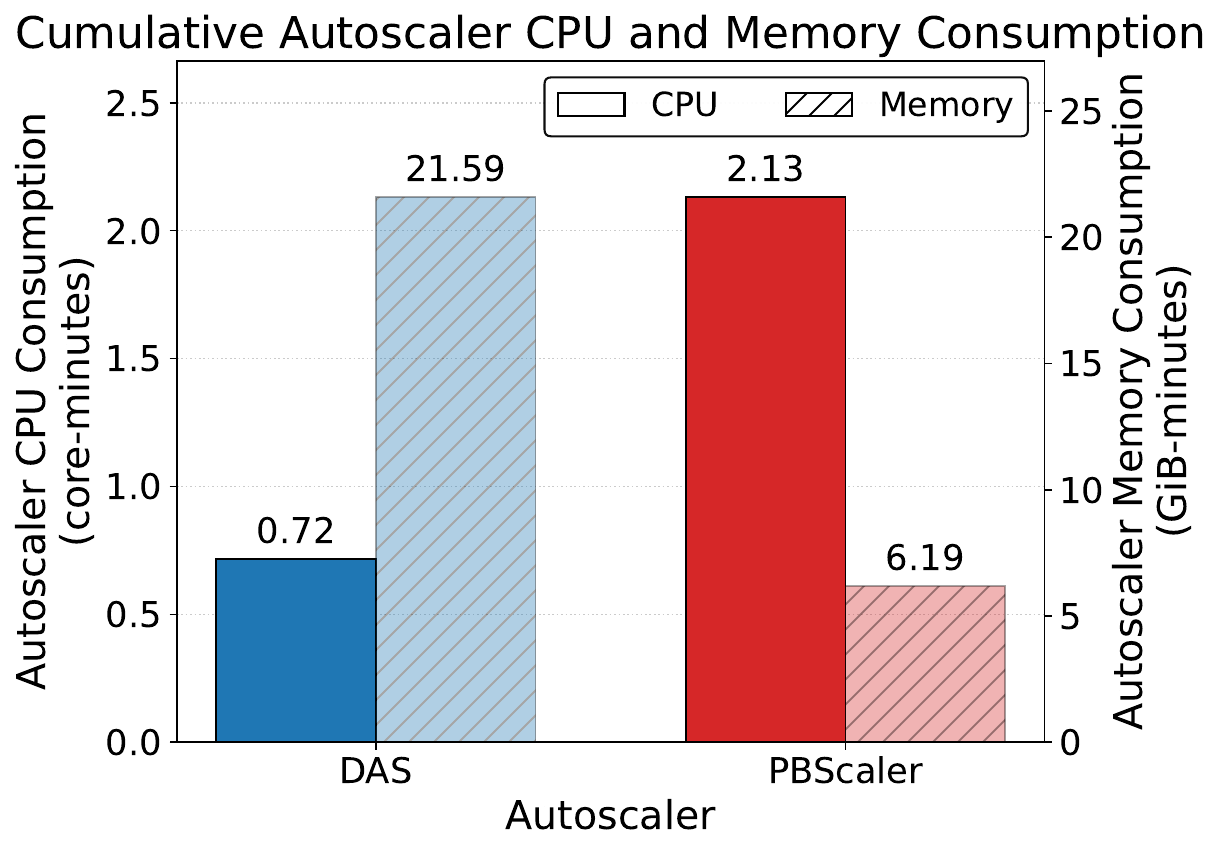}
    }

    \caption{Resource overhead of DAS and PBScaler.}\label{fig:ob_wiki_autoscaler_resource}
    
\end{figure}

\subsection{Evaluation Setup}
We created two virtual machine instances on the OpenStack cloud, each equipped with 4 vCPUs and 8 GB of memory. Our prototype was deployed on a two-node K3S cluster running Kubernetes v1.36.2, with one master node and one worker node. Istio v1.30.3 was used as the service mesh, while Prometheus v3.10.0 was deployed for metrics collection. The \tool{} server is launched as a deployment within the K3S cluster. 

As a benchmark application, we used \textbf{Online Boutique}\footnote{{https://github.com/GoogleCloudPlatform/microservices-demo}}. This application consists of 10 microservices that collectively simulate an e-commerce application, allowing users to browse products, change currencies, add items to a shopping cart, and complete purchases. To simulate realistic traffic, we utilised the popular Wiki trace~\cite{UPS09}, a real-world workload which exhibits cyclic and diurnal access patterns, to generate varying numbers of concurrent users over time. The trace was scaled to match the capacity of our cluster and to restrict the runtime to a reasonable time (i.e. 36 minutes for each experiment in this case). User behaviour was carefully modelled as a weighted mix of typical e-commerce actions to emulate realistic access patterns. Due to space limitations, we evaluate only one benchmark application and workload scenario. Nevertheless, \tool{} is inherently extensible, enabling new applications and workloads to be integrated and evaluated with minimal additional effort. Regarding autoscalers, the following three are implemented:
\begin{enumerate}

\item \textit{HPA:} Each microservice deployment is associated with a dedicated HPA configured with static target metrics. The HPA aims to maintain the specified metric at a predefined threshold. Specifically, it periodically retrieves deployment metrics and determines the desired number of replicas based on the ratio between the current and target metric values, as defined in Eq.~\ref{equ:HPA}. In our evaluation, we used CPU as the metric and configured two utilisation targets, 50\% and 80\%, resulting in two versions, HPA-50 and HPA-80, respectively.

\begin{equation}\label{equ:HPA}
\resizebox{\linewidth}{!}{$
\mathrm{desiredReplicas} =
\left\lceil 
\mathrm{currentReplicas} \times \frac{\mathrm{currentMetrics}}{\mathrm{desiredMetrics}}
\right\rceil
$}
\end{equation}

\item \textit{DAS~\cite{SGL+25}:} DAS is a decentralised autoscaler that embeds the autoscaling logic within each microservice replica. It monitors the CPU utilisation of individual replicas and partitions the utilisation space into scale-up, hold, and scale-down regions. When the utilisation falls within the scale-up or scale-down region, a probability function determines whether to add or remove a replica; otherwise, no scaling action is performed.
\item \textit{PBScaler~\cite{XWL+24}:} PBScaler is a bottleneck-aware autoscaler for microservice applications. It is triggered by SLO violations and first identifies bottleneck microservices using its TopoRank algorithm. It then employs an offline genetic algorithm, guided by a random-forest-based SLO violation predictor, to determine the replica allocation for the identified bottlenecks while minimising resource consumption and satisfying latency SLOs.
\end{enumerate} 

The selected autoscalers represent different deployment models and therefore require different integration procedures. HPA is a native Kubernetes autoscaler and can be deployed directly without modification. The original DAS implementation executes as a background process alongside each microservice replica. To integrate it into \tool{}, we reimplemented the control logic in Python, containerised the implementation, and created a deployment manifest, which was subsequently registered with the autoscaler registry. 
PBScaler is publicly available as a Python implementation but was originally executed as an external process. We containerised the original implementation, retrained the random forest model in the setup cluster, prepared a deployment manifest together with the required Kubernetes secret for cluster access, and registered it with the autoscaler registry. In both cases, algorithm-specific parameters are supplied through environment variables generated automatically from the experiment configuration.

\subsection{Evaluation criteria}
We evaluated each autoscaler using application performance, scaling, and resource metrics. Performance is measured using latency CDF, P90 latency, and SLO violation rate (requests exceeding the 500\,ms SLO). Scaling behaviour is measured by the average application replica count per minute. Application and autoscaler resource overhead are reported as cumulative CPU (core-minutes) and memory (GiB-minutes). They are computed by averaging the total resource usage per minute and summing it over the experiment.
\subsection{Computational results}
Figure~\ref{fig:performance_and_cost} summarises the performance and resource efficiency of the autoscalers. HPA-50 achieves the best latency distribution and the lowest SLO violation rate (1.00\%).
HPA-80 and DAS follow with a noticeable increase in SLO violation (4.71\% and 4.50\% respectively).
PBScaler 
exhibits the poorest latency distribution and the highest SLO violation rate (29.50\%). A closer look at the latency changes and replica counts of the application in Figure~\ref{fig:latency_replica_over_time} suggests that HPA-50 aggressively scales the application; this results in a stable latency over time~(Fig \ref{fig:ob_wiki_latency_over_time}). HPA-80 and DAS scale more conservatively, leading to slightly more latency fluctuations. In contrast, PBScaler shows frequent latency spikes despite high replica counts, indicating that scaled microservices do not relieve SLO stress.

Figure~\ref{fig:ob_wiki_autoscaler_resource} shows the resource overhead of DAS and PBScaler. Interestingly, they exhibit distinct resource consumption patterns. DAS incurs negligible CPU overhead but relatively high memory consumption, whereas PBScaler exhibits the opposite trend, consuming about 3x the CPU times while requiring less than one third of memory. Nonetheless, these results show that the resource overhead of autoscalers is non-negligible in Kubernetes clusters.

%% file: sections/05_conclusion.tex
\section{Conclusion} \label{sec:Conc}
This paper presented \tool{}, a Kubernetes-native experimental platform for reproducible and extensible evaluation of autoscaling algorithms. \tool{} deploys autoscalers inside a Kubernetes cluster, enabling a more practical evaluation of integration complexity and runtime resource overhead. The case study demonstrates the feasibility of \tool{} and its ability to provide a unified, extensible, and reproducible workflow for deploying, integrating, and benchmarking diverse Kubernetes autoscaling approaches under realistic application workloads. By evaluating both scaling performance and deployment characteristics within a common experimental framework, \tool{} facilitates fair and systematic comparison of research-based autoscalers while reducing the effort required to reproduce and extend experimental studies. Future work aims to investigate alternative Kubernetes-native mechanisms for deploying and managing autoscalers, such as Kubernetes Operators.

%% file: Ref.bib
@ARTICLE{XYA+26,
  author={Xiao, Junwei and Yang, Fan and Ai, Wei and Meng, Tao and Xie, Guoqi},
  journal={IEEE Transactions on Services Computing}, 
  title={PAHPA: Revolutionizing Kubernetes Autoscaling With Integrated Predictive Analytics and Real-Time Monitoring}, 
  year={2026},
  volume={19},
  number={3},
  pages={2343-2355},
  doi={10.1109/TSC.2026.3685325}}

@article{CRB+11,
author = {Calheiros, Rodrigo N. and Ranjan, Rajiv and Beloglazov, Anton and De Rose, César A. F. and Buyya, Rajkumar},
title = {CloudSim: a toolkit for modeling and simulation of cloud computing environments and evaluation of resource provisioning algorithms},
journal = {Software: Practice and Experience},
volume = {41},
number = {1},
pages = {23-50},
doi = {https://doi.org/10.1002/spe.995},
url = {},
eprint = {https://onlinelibrary.wiley.com/doi/pdf/10.1002/spe.995},
year = {2011}
}

@article{ATTG21,
title = {AutoScaleSim: A simulation toolkit for auto-scaling Web applications in clouds},
journal = {Simulation Modelling Practice and Theory},
volume = {108},
pages = {102245},
year = {2021},
issn = {1569-190X},
doi = {https://doi.org/10.1016/j.simpat.2020.102245},
url = {},
author = {Mohammad S. Aslanpour and Adel N. Toosi and Javid Taheri and Raj Gaire}
}

@ARTICLE{DLC+25,
  author={Ding, Yu and Li, Chenhao and Cai, Zhengong and Wang, Xinghao and Yang, Bowei},
  journal={IEEE Access}, 
  title={A Dynamic Interval Auto-Scaling Optimization Method Based on Informer Time Series Prediction}, 
  year={2025},
  volume={13},
  number={},
  pages={14572-14583},
  doi={10.1109/ACCESS.2024.3513564}}

@ARTICLE{WZL+24,
  author={Wang, Ziliang and Zhu, Shiyi and Li, Jianguo and Jiang, Wei and Ramakrishnan, K. K. and Yan, Meng and Zhang, Xiaohong and Liu, Alex X.},
  journal={IEEE/ACM Transactions on Networking}, 
  title={DeepScaling: Autoscaling Microservices With Stable CPU Utilization for Large Scale Production Cloud Systems}, 
  year={2024},
  volume={32},
  number={5},
  pages={3961-3976},
  doi={10.1109/TNET.2024.3400953}}

@INPROCEEDINGS{CLG+24,
  author={Cao, Hongchen and Liu, Xinrui and Guo, Hengquan and He, Jingzhu and Liu, Xin},
  booktitle={2024 IEEE International Conference on Web Services (ICWS)}, 
  title={QueueFlower: Orchestrating Microservice Workflows via Dynamic Queue Balancing}, 
  year={2024},
  volume={},
  number={},
  pages={1293-1299},
  doi={10.1109/ICWS62655.2024.00155}}

@ARTICLE{XWL+24,
  author={Xie, Shuaiyu and Wang, Jian and Li, Bing and Zhang, Zekun and Li, Duantengchuan and Hung, Patrick C. K.},
  journal={IEEE Transactions on Services Computing}, 
  title={PBScaler: A Bottleneck-Aware Autoscaling Framework for Microservice-Based Applications}, 
  year={2024},
  volume={17},
  number={2},
  pages={604-616},
  doi={10.1109/TSC.2024.3376202}}

@article{Carrion22,
author = {Carri{\'o}n, Carmen},
title = {Kubernetes Scheduling: Taxonomy, Ongoing Issues and Challenges},
year = {2022},
issue_date = {July 2023},
publisher = {Association for Computing Machinery},
address = {New York, NY, USA},
volume = {55},
number = {7},
issn = {0360-0300},
url = {},
doi = {10.1145/3539606},
journal = {ACM Comput. Surv.},
month = dec,
articleno = {138},
numpages = {37}
}

@INPROCEEDINGS{TWPY21,
  author={Tong, Jingwan and Wei, Mingchang and Pan, Maolin and Yu, Yang},
  booktitle={2021 IEEE International Conference on Web Services (ICWS)}, 
  title={A Holistic Auto-Scaling Algorithm for Multi-Service Applications Based on Balanced Queuing Network}, 
  year={2021},
  volume={},
  number={},
  pages={531-540},
  doi={10.1109/ICWS53863.2021.00074}}

@Article{MWK+23,
AUTHOR = {Mondal, Subrota Kumar and Wu, Xiaohai and Kabir, Hussain Mohammed Dipu and Dai, Hong-Ning and Ni, Kan and Yuan, Honggang and Wang, Ting},
TITLE = {Toward Optimal Load Prediction and Customizable Autoscaling Scheme for Kubernetes},
JOURNAL = {Mathematics},
VOLUME = {11},
YEAR = {2023},
NUMBER = {12},
ARTICLE-NUMBER = {2675},
URL = {},
ISSN = {2227-7390},
DOI = {10.3390/math11122675}
}

@article{ZTL+26,
  title={Load-aware predictive auto-scaling framework for cloud environments},
  author={Zanussi, Luca and Tessera, Daniele and Massari, Luisa and Bermejo, Belen and Juiz, Carlos and Calzarossa, Mariacarla},
  journal={Cluster Computing},
  volume={29},
  number={3},
  pages={197},
  year={2026},
  publisher={Springer}
}

@INPROCEEDINGS{DT25,
  author={Dashtbani, Majid and Tahvildari, Ladan},
  booktitle={2025 IEEE International Conference on Collaborative Advances in Software and COmputiNg (CASCON)}, 
  title={Key Considerations for Auto-Scaling: Lessons from Benchmark Microservices}, 
  year={2025},
  volume={},
  number={},
  pages={118-123},
  doi={10.1109/CASCON66301.2025.00034}}

@INPROCEEDINGS{TVK22,
  author={Tran, Minh-Ngoc and Vu, Dinh-Dai and Kim, Younghan},
  booktitle={2022 Thirteenth International Conference on Ubiquitous and Future Networks (ICUFN)}, 
  title={A Survey of Autoscaling in Kubernetes}, 
  year={2022},
  volume={},
  number={},
  pages={263-265},
  doi={10.1109/ICUFN55119.2022.9829572}}

@Article{NYK+20,
AUTHOR = {Nguyen, Thanh-Tung and Yeom, Yu-Jin and Kim, Taehong and Park, Dae-Heon and Kim, Sehan},
TITLE = {Horizontal Pod Autoscaling in Kubernetes for Elastic Container Orchestration},
JOURNAL = {Sensors},
VOLUME = {20},
YEAR = {2020},
NUMBER = {16},
ARTICLE-NUMBER = {4621},
URL ={},
PubMedID = {32824508},
ISSN = {1424-8220},
DOI = {10.3390/s20164621}
}

@book{LC26,
  title={Kubernetes in Action: Deploying and managing containers and cloud-native applications},
  author={Luk{\v{s}}a, Marko and Conner, Kevin},
  year={2026},
  publisher={Simon and Schuster}
}

@article{prometheus,
  title={Prometheus: A $\{$Next-Generation$\}$ monitoring system (talk)},
  author={Rabenstein, Bj{\"o}rn and Volz, Julius},
  year={2015}
}

@inproceedings{SEK+23,
author = {Straesser, Martin and Eismann, Simon and von Kistowski, J\'{o}akim and Bauer, Andr\'{e} and Kounev, Samuel},
title = {Autoscaler Evaluation and Configuration: A Practitioner's Guideline},
year = {2023},
isbn = {9798400700682},
publisher = {Association for Computing Machinery},
address = {New York, NY, USA},
url = {},
doi = {10.1145/3578244.3583721},
booktitle = {Proceedings of the 2023 ACM/SPEC International Conference on Performance Engineering},
pages = {31–41},
numpages = {11},
location = {Coimbra, Portugal},
series = {ICPE '23}
}

@inproceedings{XWL+26,
author = {Xie, Shuaiyu and Luo, Yang and Tan, Yuzhen and Yong, Yunqing and Huang, Xiaosong and Wang, Jian and Li, Bing},
title = {ScalerEval: Automated and Consistent Evaluation Testbed for Auto-scalers in Microservices},
year = {2026},
isbn = {9798400726361},
publisher = {Association for Computing Machinery},
address = {New York, NY, USA},
url = {},
doi = {10.1145/3803437.3806415},
booktitle = {Proceedings of the 34th ACM International Conference on the Foundations of Software Engineering},
pages = {177–181},
numpages = {5},
location = {Concordia University, Montreal, QC, Canada},
series = {FSE Companion '26}
}

@inproceedings{SGK+22,
author = {Straesser, Martin and Grohmann, Johannes and von Kistowski, J\'{o}akim and Eismann, Simon and Bauer, Andr\'{e} and Kounev, Samuel},
title = {Why Is It Not Solved Yet? Challenges for Production-Ready Autoscaling},
year = {2022},
isbn = {9781450391436},
publisher = {Association for Computing Machinery},
address = {New York, NY, USA},
url = {},
doi = {10.1145/3489525.3511680},
booktitle = {Proceedings of the 2022 ACM/SPEC on International Conference on Performance Engineering},
pages = {105–115},
numpages = {11},
location = {Beijing, China},
series = {ICPE '22}
}

@article{DNK23,
title = {Auto-scaling techniques in container-based cloud and edge/fog computing: Taxonomy and survey},
journal = {Computer Communications},
volume = {209},
pages = {120-150},
year = {2023},
issn = {0140-3664},
doi = {https://doi.org/10.1016/j.comcom.2023.06.010},
url = {},
author = {Javad Dogani and Reza Namvar and Farshad Khunjush}
}

@article{JJ25,
title = {Autoscaling techniques in cloud-native computing: A comprehensive survey},
journal = {Computer Science Review},
volume = {58},
pages = {100791},
year = {2025},
issn = {1574-0137},
doi = {https://doi.org/10.1016/j.cosrev.2025.100791},
url = {},
author = {Byeonghui Jeong and Young-Sik Jeong}
}

@INPROCEEDINGS{ATWS24,
  author={Ahmad, Hussain and Treude, Christoph and Wagner, Markus and Szabo, Claudia},
  booktitle={2024 IEEE 21st International Conference on Software Architecture (ICSA)}, 
  title={Smart HPA: A Resource-Efficient Horizontal Pod Auto-Scaler for Microservice Architectures}, 
  year={2024},
  volume={},
  number={},
  pages={46-57},
  doi={10.1109/ICSA59870.2024.00013}}

@article{RCPN20,
title = {Geo-distributed efficient deployment of containers with Kubernetes},
journal = {Computer Communications},
volume = {159},
pages = {161-174},
year = {2020},
issn = {0140-3664},
doi = {https://doi.org/10.1016/j.comcom.2020.04.061},
url = {},
author = {Fabiana Rossi and Valeria Cardellini and Francesco {Lo Presti} and Matteo Nardelli}
}

@article{UPS09,
title = {Wikipedia workload analysis for decentralized hosting},
journal = {Computer Networks},
volume = {53},
number = {11},
pages = {1830-1845},
year = {2009},
issn = {1389-1286},
doi = {https://doi.org/10.1016/j.comnet.2009.02.019},
url = {},
author = {Guido Urdaneta and Guillaume Pierre and Maarten {van Steen}}
}

@article{SGL+25,
title = {Trust your local scaler: A continuous, decentralized approach to autoscaling},
journal = {Performance Evaluation},
volume = {167},
pages = {102452},
year = {2025},
issn = {0166-5316},
doi = {https://doi.org/10.1016/j.peva.2024.102452},
url = {},
author = {Martin Straesser and Stefan Geissler and Stanislav Lange and Lukas Kilian Schumann and Tobias Hossfeld and Samuel Kounev}
}

@article{ullah2023orchestration,
  title={Orchestration in the cloud-to-things compute continuum: taxonomy, survey and future directions},
  author={Ullah, Amjad and Kiss, Tamas and Kov{\'a}cs, J{\'o}zsef and Tusa, Francesco and Deslauriers, James and Dagdeviren, Huseyin and Arjun, Resmi and Hamzeh, Hamed},
  journal={Journal of Cloud Computing},
  volume={12},
  number={1},
  pages={1--29},
  year={2023},
  publisher={Springer}
}

@article{ullah2018control,
  title={A control theoretical view of cloud elasticity: taxonomy, survey and challenges},
  author={Ullah, Amjad and Li, Jingpeng and Shen, Yindong and Hussain, Amir},
  journal={Cluster Computing},
  volume={21},
  number={4},
  pages={1735--1764},
  year={2018},
  publisher={Springer}
}
